\documentclass[doublecol]{epl2} 

\usepackage{amsmath}  
\usepackage{amsfonts} 
\usepackage{graphicx} 
\usepackage{color}
\usepackage{lipsum}
\usepackage{ulem}
\usepackage{subfigure}

\newcommand {\ggtau} {g^{(2)}(\tau)}
\newcommand {\gtau} {g^{(1)}(\tau)}

\title{Temporal coherences of atomic chaotic light sources: the Siegert relation and beyond}
\shorttitle{Temporal correlations for chaotic light} 

\author{M. Morisse\inst{1} \and S. Joshi\inst{2} \and J. Mika\inst{2} \and J. C. C. Capella\inst{1} \and R. Kaiser\inst{1} \and R. Bachelard\inst{3} \and L. Slodi\v{c}ka\inst{2} \and M. Hugbart\inst{1}}
\shortauthor{M. Morisse \etal}

\institute{                    
  \inst{1} Universit\'e C\^ote d'Azur, CNRS, INPHYNI, France\\
  \inst{2} Department of Optics, Palacký University, 17. listopadu 12, 771 46 Olomouc, Czech Republic\\
  \inst{3} Departamento de F\'{\i}sica, Universidade Federal de S\~{a}o Carlos, Rodovia Washington Lu\'{\i}s, km 235 - SP-310, 13565-905 S\~{a}o Carlos, SP, Brazil
}

\abstract{
Light is characterized by its electric field, yet quantum optics has revealed the importance of monitoring photon-photon correlations at all orders. We here present a comparative study of two experimental setups, composed of cold and warm Rubidium atoms, respectively, which allow us to probe and compare photon correlations up to the fourth order. The former operates in the quantum regime where spontaneous emission dominates, whereas the latter exhibits a temperature-limited coherence time. While both setups present almost-chaotic light statistics, we discuss how the access to different orders of photon correlations allows one to better characterize the mechanisms responsible for deviations from those statistics.
}

\begin{document}

\maketitle

\section{Introduction}

Quantum mechanics was applied to optics, with Roy J. Glauber as one of the pioneers of what became the field of quantum optics~\cite{Glauber:1963a}. He introduced a deeper description of the statistics of light and a precise definition of optical coherences. Glauber's approach is based on the temporal and spatial correlation functions of the light field, with a complete description of the coherence properties of a light source requiring the measurements of these correlations at all orders. Practically, the first orders are usually used to distinguish different types of sources, either classical or quantum, such as thermal sources, lasers or single photon sources.

Among the light sources sharing common statistical properties, the most common one found in nature is probably those whose field exhibits Gaussian statistics with zero mean. In the literature, those sources are referred to by various names such as chaotic, thermal or Gaussian sources. Although their statistical features are the same, the physical processes behind them are not necessarily the same. On the one hand, thermal source usually refers to thermal radiation and thus light coming from a black body in thermal equilibrium such as, to a good approximation, the light emitted by the stars or by the heated filament of a light bulb. On the other hand, 'chaotic' emphasizes the underlying randomness of the emission process at the microscopic level, with light coming from a large number of uncorrelated emitters. Such mechanism emerges, for instance, from collisions in a gas-discharge lamp, from motion of disordered defects in the case of a rotating diffuser (it is then also called pseudo-thermal light)~\cite{Martienssen_1964}, or to temperature (and the subsequent Doppler broadening) and/or spontaneous emission for the light scattered by a disordered cloud of cold atoms~\cite{Lassegues2023}.

The Gaussian nature of the electric field and its zero mean can be witnessed through different kinds of measurements. More specifically, the associated intensity probability distribution follows an exponential decay, $P(I) = \exp{(-I/\langle I \rangle})/\langle I \rangle$~\cite{Goodman:1976}, whose moments are given by: $\langle I^n \rangle/\langle I \rangle^n = n!$\,\cite{Glauber:1963b}. These moments describe the temporal intensity correlation function of order $n\ge 2$ at zero delay, $g^{(n)}(0)$, yet a thorough characterization of the light requires examining the relation between the temporal correlation function of the electric field at all orders and for all delays, $g^{(n)}
(\tau)$.

The implementation and control of nontrivial photon correlations is also of paramount interest in quantum physics. Generated from highly nonlinear interactions, their precise characterization often requires the measurements of higher moments of the generated light fields~\cite{Lemieux1999,titulaer1965correlation,sperling2015uncovering,laiho2022measuring}. For some quantum systems such as ultracold atoms~\cite{dall2013ideal}, such measurements are increasingly common, with the broad objective of studying quantum many-body phenomena. Yet systematic measurements for quantum light sources have been precluded by the stringent experimental requirements to observe photon correlation functions $g^{(n)}$ with $n>2$, typically due to the low photon flux. In order to provide a more precise characterization of the quantum light states generated, and of the underlying processes, a strong effort is being realized over multiple experimental platforms to access higher-order photon correlations~\cite{laiho2022measuring,liang2018observation,patil2022measuring,assmann2009higher}.

In this paper, we focus on the temporal coherence and the measurement of the corresponding correlation functions in two atomic setups producing near-chaotic light. For such sources and in the stationary regime (an assumption followed throughout this work), one can show that the higher order temporal correlation functions are simply determined by the light spectrum or, equivalently, by its Fourier transform, the electric field temporal correlation function\,\cite{Glauber:1963b}. In the following, the electric field autocorrelation and the intensity correlation functions at different orders are connected through the Siegert relation~\cite{Siegert:1943} and its higher-order generalization in the framework of classical theory, but one can show that quantum theory yields the same prediction for chaotic light\,\cite{Loudon:book}. The first atomic setup, operating with cold atoms, provides access to the simultaneous measurements of field-field and photon-photon correlation functions, allowing for their direct comparison without any shot-to-shot fluctuations; the second setup, based on a warm atomic vapor, is able to monitor three- and four-photon correlations, thanks to a multiplexed array of single-photon detectors. These measurements allow us to discuss the validity of the Siegert relation and its higher-order generalizations in the context of these two atomic sources of near-chaotic light.


\section{First and second order correlation functions: the Siegert relation with cold atoms}

\subsection{First order correlation function}

A traditional method to characterize light coherence is the temporal autocorrelation of the complex electric field $E(t)$:
\begin{equation}\label{eq:g1}
    g^{(1)}(\tau) = \frac{\langle E^\star(t) E(t + \tau) \rangle}{\langle I(t) \rangle},
\end{equation}
with $I(t) = E^\star(t)E (t)$ the intensity, and $\langle . \rangle$ the ensemble average over a stationary process (it corresponds to the temporal average for an ergodic system, or the expectation value for quantum processes). This autocorrelation is commonly measured with interferometric setups, but other techniques exist such as spectroscopic measurements, where the Fourier transform of $\gtau$ provides the light spectrum, or the beatnote technique. Next-order correlations correspond to the (temporal) intensity autocorrelation function, defined as:
\begin{equation}
    g^{(2)}(\tau) = \frac{\langle I(t)I(t + \tau) \rangle}{\langle I(t) \rangle^2}. \label{eq:g2}
\end{equation}

\begin{figure}[h]
	\centering
	\includegraphics[width=0.9\columnwidth]{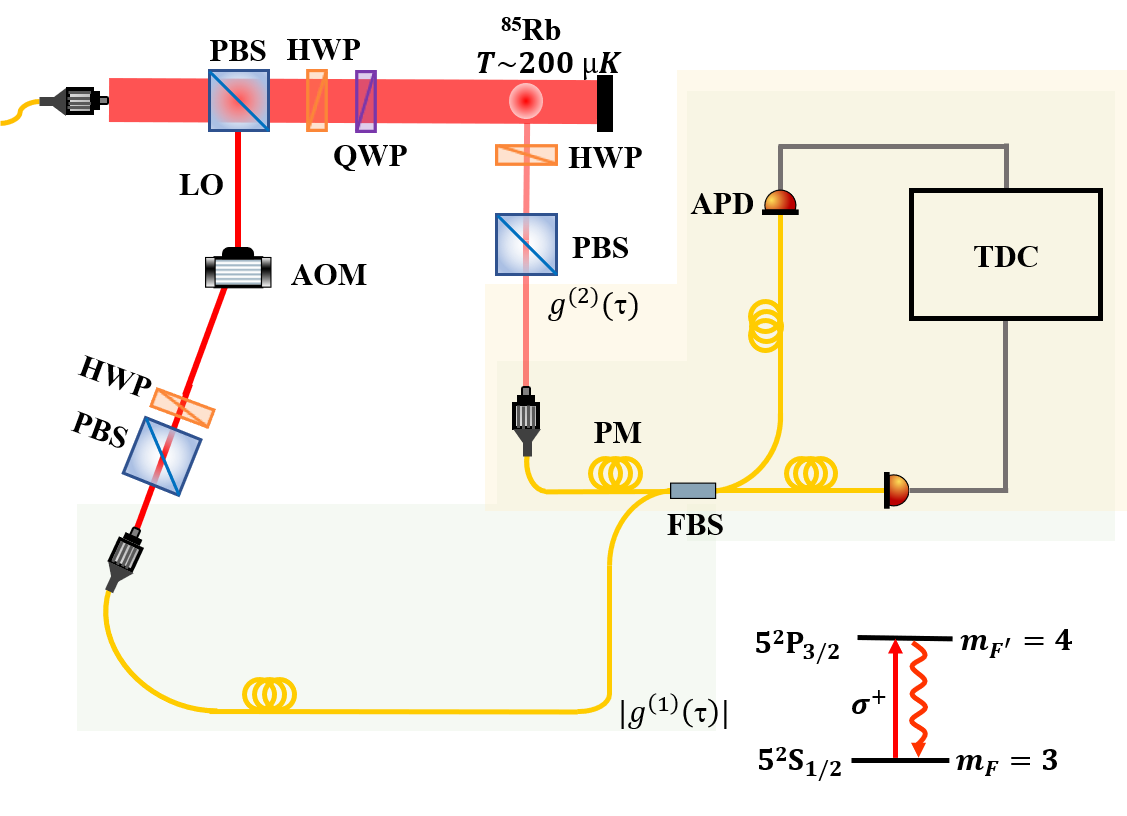}
	\caption{Experimental setup to simultaneously measure $\gtau$ and $\ggtau$ of the light scattered by cold atoms illuminating by a circularly polarized laser beam. The scattered light is collected by a polarization-maintaining (PM) single-mode fiber. The light is then split to illuminate two avalanche photodiodes (APDs). Each photon arrival is time-tagged by a time-to-digital converter (TDC). Finally, a local oscillator (LO), derived from the incident laser and frequency-shifted by an acousto-optical modulator (AOM), is injected in the second input of the fibered beam splitter (FBS). Waveplates (QWP: quarter waveplate, HWP: half waveplate) and polarizing beam splitter (PBS) are used to select the polarization and to maximize interference contrast when the scattered light interferes with the LO.}
	\label{fig:Setup_CA}
\end{figure}

The first platform presented in this paper is based on cold atoms. Depicted in Fig.\,\ref{fig:Setup_CA}, it provides simultaneous access to both autocorrelation functions. The setup has been described in detail in Refs.\,\cite{Ortiz_2019, Lassegues_2022}: The scattering medium corresponds to $^{85}$Rb atoms at about 200\,$\mu$K released from a magneto-optical trap. The atoms are uniformly illuminated by a circularly polarized laser beam at resonance with the $\lvert 3\rangle \rightarrow \lvert 4'\rangle$ hyperfine transition of the D2 line. The optical thickness is lower than 1, so multiple scattering can be neglected.

The scattered light is collected at an angle different from zero from the probe beam axis using the PM fiber. The latter is connected to a splitter, whose outputs illuminate two single photon counter detectors, thus implementing a standard ``Hanbury Brown and Twiss" (HBT) setup~\cite{HBT:1956a}. The two detectors allow overcoming both their deadtime, of the order of a few tens of nanoseconds for this setup and which would prevent the detection of two photons by one detector with a delay smaller than the deadtime, and the impact of the detector afterpulsing. However, the splitter and the two detectors are not fundamentally required, neither for classical light nor quantum light. Indeed, experiments performed with only one detector have also reported $\ggtau$ measurements revealing the quantum nature of light\,\cite{Steudle2012}. Finally, while the arrival of each photon is time-tagged by the TDC, the second input of the FBS is used to inject a local oscillator (LO) derived from the incident laser and frequency shifted with an acousto-optical modulator by $\omega_\mathrm{BN}$. 

\begin{figure}[h]
	\centering
	\includegraphics[width=0.9\columnwidth]{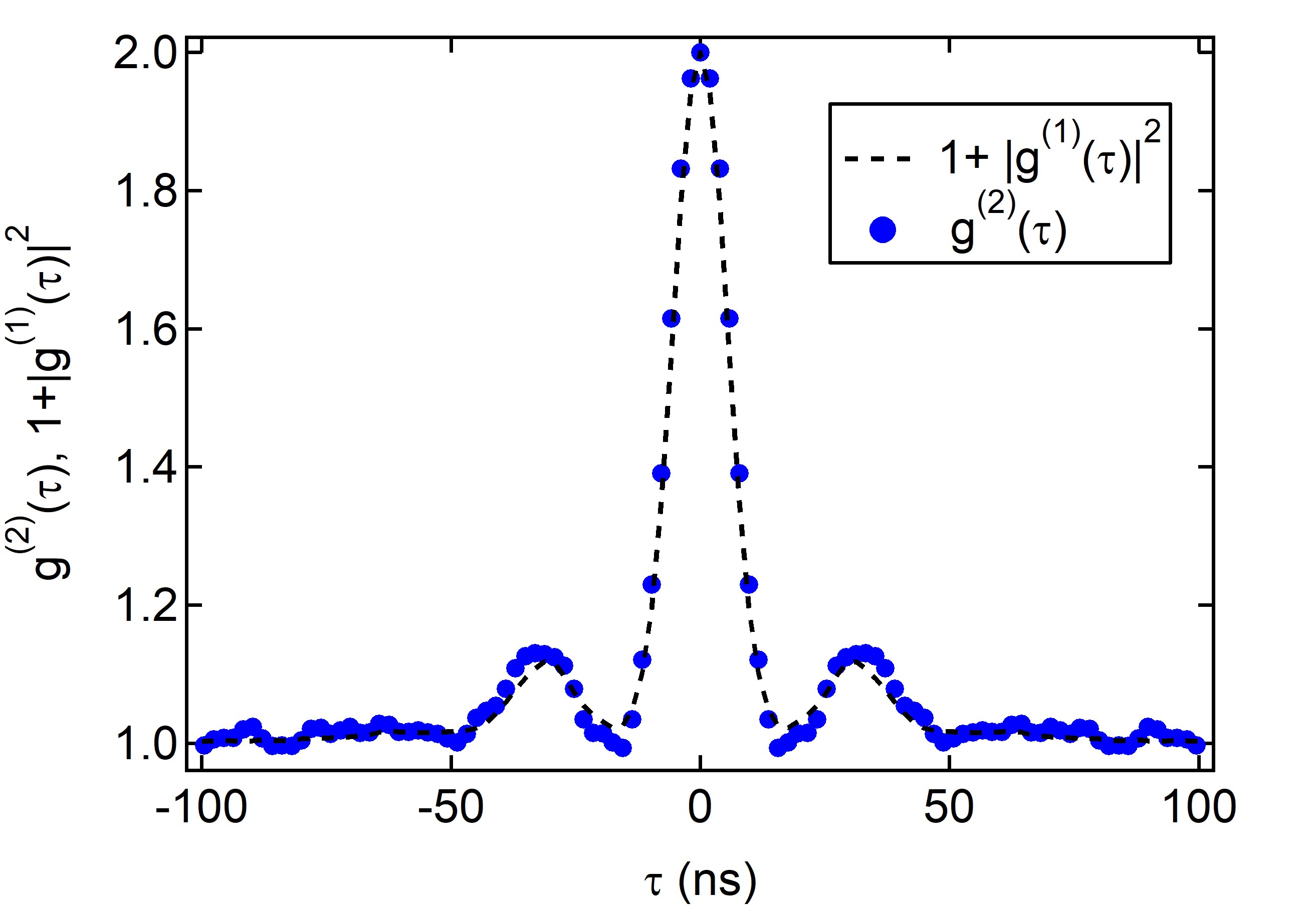}
	\caption{Experimental signals of $1+|\gtau|^2$ and $\ggtau$ for the light scattered by a cold atomic cloud illuminated with a laser intensity corresponding to $s_0 \approx 60$. The overlap of the two curves is a manifestation of the validity of the Siegert relation.}
	\label{fig:g2_Siegert_CA}
\end{figure}

The $\gtau$ and $\ggtau$ functions are extracted through the measurement of the temporal correlation of the intensity of the beat note $g_\mathrm{BN}{}^{(2)}(\tau)$ between the scattered light and the LO\,\cite{Ferreira2020}. Thanks to the AOM, the two quantities $\gtau$ and $\ggtau$ are separated in Fourier space, with $\ggtau$ centered around the DC value and $\gtau$ centered at $\omega_\mathrm{BN}$. Then, they are separately computed back in the temporal space, the oscillatory component at frequency $\omega_\mathrm{BN}$ being eliminated. Fig.\,\ref{fig:g2_Siegert_CA} presents an example of the autocorrelation functions obtained with the cold atom experiment. The atoms are illuminated at resonance with a saturation parameter $s_0\approx 60$, so the scattering is mainly inelastic: Its spectrum exhibits the well-known Mollow triplet~\cite{Mollow_1969}, characterized by a carrier and two sidebands. The beating between these spectral components corresponds, in the time space, to Rabi oscillations. As it can be observed in Fig.\,\ref{fig:g2_Siegert_CA}, the two autocorrelation functions are connected by the relation $\ggtau=1+|\gtau|^2$ which, as we shall now see, results from the Gaussian nature of the electric field. 


\subsection{Connecting the field correlation functions: the Siegert relation}

Assuming a quasi-monochromatic field at optical frequency, the total radiated complex electric field can be written as $E(t) = E_0(t) e^{-i \omega t}$, 
while the complex amplitude $E_0(t)$ is a slowly fluctuating process. 
We hereafter adopt the mathematical formalism detailed in Ref.\,\cite{Goodman:StatOptics}, using the complex notation: 
its real part $\Re{[E(t)]}$ denotes the classical electric field, while the imaginary part $\Im{[E(t)]}$ corresponds to the Hilbert transform of the real part\,\cite{Goodman:StatOptics}. The statistics of $E(t)$ are thus completely determined by those of $\Re{[E(t)]}$. One can also show that the autocorrelations $\langle \Re{[E(t)]} \Re{[E(t+\tau)]} \rangle$ and $\langle \Im{[E(t)]} \Im{[E(t+\tau)]} \rangle$ are equal, while the cross correlations $\langle \Re{[E(t)]} \Im{[E(t+\tau)]} \rangle$ and $\langle \Im{[E(t)]} \Re{[E(t+\tau)]} \rangle$ have opposite signs.

We now assume that the light field can be treated as a Gaussian random process, as is the case, for example, for light scattered by a large number of statically independent particles. This means that $\Re{[E(t)]}$ is a Gaussian random process as well as the imaginary part, the Gaussian statistics being preserved under linear operations such as the Hilbert transformation, and the complex electric field is thus a complex Gaussian random process -- see definition in Ref.\,\cite{mandel1995optical} based on the distribution characterisation of $n$-fold joint probabilities. Furthermore, we assume that the electric field amplitude has a zero mean, $\langle E_0(t)\rangle = 0$, and this class of random variables is called circular complex Gaussian variable\,\cite{Goodman:StatOptics}.
Applying Isserlis' theorem\,\cite{Isserlis_1918}, also known as the Wick's probability theorem in quantum field theory\,\cite{wick1950evaluation}, the expression for intensity autocorrelation \eqref{eq:g2} leads to the Siegert relation\,\cite{Siegert:1943}:
\begin{equation}\label{eq:Siegert_simple}
    g^{(2)}(\tau) = 1+\lvert g^{(1)}(\tau)\rvert^2.
\end{equation}
An additional term, $|\langle E(t)E(t+\tau)\rangle|^2/\langle I(t)\rangle^2$, is sometimes added, as in Ref.\,\cite{Lemieux1999}. This term is actually equal to zero due to the properties of the autocorrelations and cross correlations of the real and imaginary parts of the electric field described above\,\cite{Goodman:StatOptics}.

The validity of this relation for the light scattered in the cold atom setup can be appreciated in Fig.\,\ref{fig:g2_Siegert_CA}. The procedure to extract simultaneously $g^{(1)}$ and $g^{(2)}$, and compare them, is detailed in Ref.\,\cite{Ferreira2020}. In particular, once $\gtau$ and $\ggtau$ are derived from $g_\mathrm{BN}$, $\gtau$ is normalised to 1 at zero delay, as well as the $\ggtau$ contrast, $g^{(2)}(0)-g^{(2)}(\infty)$. While this normalization is set for $\gtau$ from its very definition [see Eq.\,(\ref{eq:g1})], it is not the case for $\ggtau$. Indeed, already for classical fields, $g^{(2)}(0)$ can reach values greater than two, going from Poissonian to superbunched light\,\cite{Ficek_2005,Marconi_2018}. For quantum light, $\ggtau$ can take any value greater or equal to 0. In the cold atom experiment, the validity of the normalisation to 2 has been checked by measuring separately the $\ggtau$ function of the scattered light without the LO, when no normalisation is needed\,\cite{Lassegues_2022}. 

This setup has been first implemented with low saturation parameter\,\cite{Ferreira2020}, the Gaussian statistics of the electric field coming from the atomic velocity distribution. The very good overlap between the $\gtau$ and $\ggtau$ correlation functions in Fig.\,\ref{fig:g2_Siegert_CA} demonstrates the validity of the Siegert relation for large saturation parameter, where one needs to take into account the quantum nature of the scatterers. Indeed, for inelastic scattering, Gaussian statistics for the electric field and zero expectation value are achieved due to the large number of cold atoms and to the (quantum) randomness of spontaneous emission\,\cite{Lassegues2023}. This latter effect is also responsible for the coherence time of a few tens of nanoseconds for the scattered light, which corresponds to the excited state lifetime. Indeed, the sufficiently low temperature of the atoms allows being sensitive to quantum aspects such as spontaneous emission, which would otherwise be hidden for larger temperature, such as in hot vapors.

The validity of the Siegert relation has been reported in several scattering media such as polystyrene spheres in aqueous suspension\,\cite{Wolf1988} or cold atoms\,\cite{Ferreira2020,Lassegues_2022, Ferioli2023}. Yet its violation is also possible, indicating that one of the assumptions underlying its derivation is not valid: either the mean expectation value of the electric field complex amplitude is non zero, and/or the statistics are not Gaussian. This is the case when the scatterers are few in numbers, or are correlated\,\cite{Voigt1994,Kovalenko_2023,Ferioli2023}. A non-zero average value can also stem from a coherent component of the scattered field, such as in the forward direction where interference occur. For such non-Gaussian field, higher order correlation functions may provide access to additional information not contained in $\ggtau$\cite{Lemieux1999}.

\section{Higher order correlations with warm atoms}

In this section, we present the measurements of the light coherence from a warm atom vapor, and achieve the characterization of the photon correlation functions up to the fourth order. Using a near-single-mode detection scheme~\cite{mika2018generation}, and due to the temperature-induced phase randomization of the emission, the sample is expected to produce chaotic light: The measured correlations functions are indeed in close agreement with the theoretical prediction for Gaussian statistics~\cite{Isserlis_1918,wick1950evaluation}.  
\begin{figure}[t!]
	\centering	\includegraphics[width=\columnwidth]{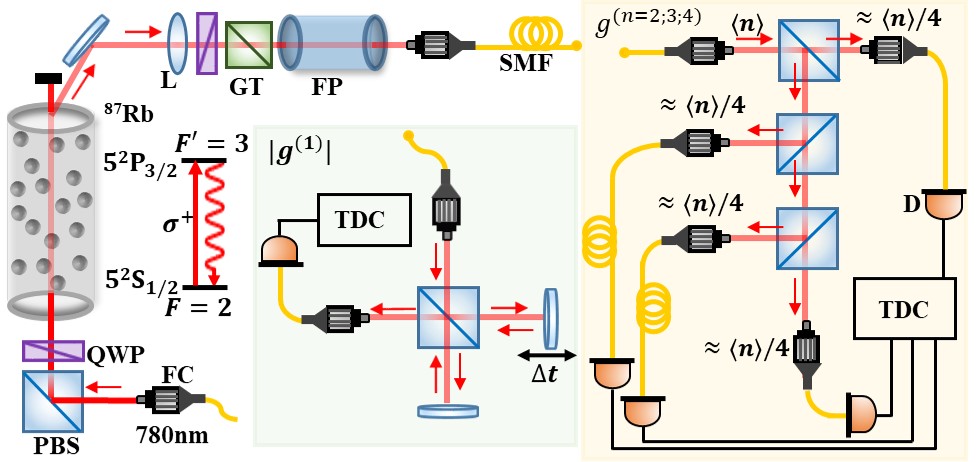}
	\caption{Schematic of the experimental setup for the measurement of first- and higher-order correlations of the light scattered from a warm $^{87}\text{Rb}$ vapor. The scattering of a single laser beam at 780~nm by atoms in an approximate two-level configuration and in the weak saturation limit results in the generation of chaotic light. The first degree of coherence $|g^{(1)}(\tau)|$ is evaluated by employing Michelson interferometer with various time delays. The higher-order photon correlations $g^{(n)}$ for $n=2,3,4$ are accessed using a spatially multiplexed array of single-photon detectors. The optical components were chosen to guarantee the single-mode detection of chaotic light. SMF - single mode fiber; L - lens; GT - Glan-Thomson polarizer; FP - Fabry-Perot resonator.
 }
	\label{fig:setup_vapor}
\end{figure}

This experimental setup is based on the observation of scattered light from a warm vapor of $^{87}\text{Rb}$ atoms. An approximation of a two-level interaction limit is achieved by implementation of weak resonant excitation of 
$5{\rm S}_{1/2}({\rm F}=2) \leftrightarrow 5{\rm P}_{3/2}({\rm F'}=3)$ transition using a circularly polarized $780$~nm laser beam. The vapor cell was set to a temperature of $64^{\circ}$C and the scattered light is observed at a small angle of about $1.7\,^{\circ}$, which significantly reduces the Doppler broadening. As illustrated in Fig.~\ref{fig:setup_vapor}, the interaction volume close to the output of the vapor cell is determined 
by the overlap between excitation and observation optical spatial modes. The estimated saturation parameter for the measured excitation laser power of $P=300\,\mu$W is $s_0=0.14$. A sequence of polarization optical components and Fabry-Pérot frequency filter in the observation optical mode is used to suppress the contribution of photons from Raman transitions to the $5{\rm S}_{1/2}({\rm F}=1)$ manifold to the detection mode. The sufficiently large linewidth, with a full width at half maximum (FWHM) of~$0.9$ GHz and free spectral range of $30$~GHz of the Fabry-Pérot filter, has been selected such that it does not affect the observable spectral distributions of photons scattered from the chosen two-level transition even for multiply-scattered photons~\cite{dussaux2016temporal,Eloy_2018}. A single spatial detection mode is guaranteed by coupling the scattered light to a PM single-mode optical fiber. The coherence properties are first analyzed in the setup by including a Michelson interferometer, providing access to the modulus of first-order correlation, $|\gtau|$. The correlation functions of order $n\geq 2$ are accessed in the low mean photon flux regime by modifying the HBT setup into a spatially multiplexed array of single-photon detectors~\cite{oppel2012superresolving}. The observation of near-chaotic photon correlations up to the fourth order is achieved by relatively high generated and collected photon flux from the warm atomic ensemble, $\langle n\rangle= (45.8\pm 0.5)\times 10^{-4}$ photons at the chosen $1.94$~ns time bin, and thanks to the long coherence time of forward scattered photons as compared with the time resolution of the detectors.  Alternatively, temporal or spatio-temporal multiplexing schemes~\cite{avenhaus2010accessing,qi2018multiphoton}, or superconducting nanowire detectors with photon-number resolving capabilities can be employed~\cite{gol2001picosecond}. 

\begin{figure}[t!]
	\centering	\includegraphics[width=\columnwidth]{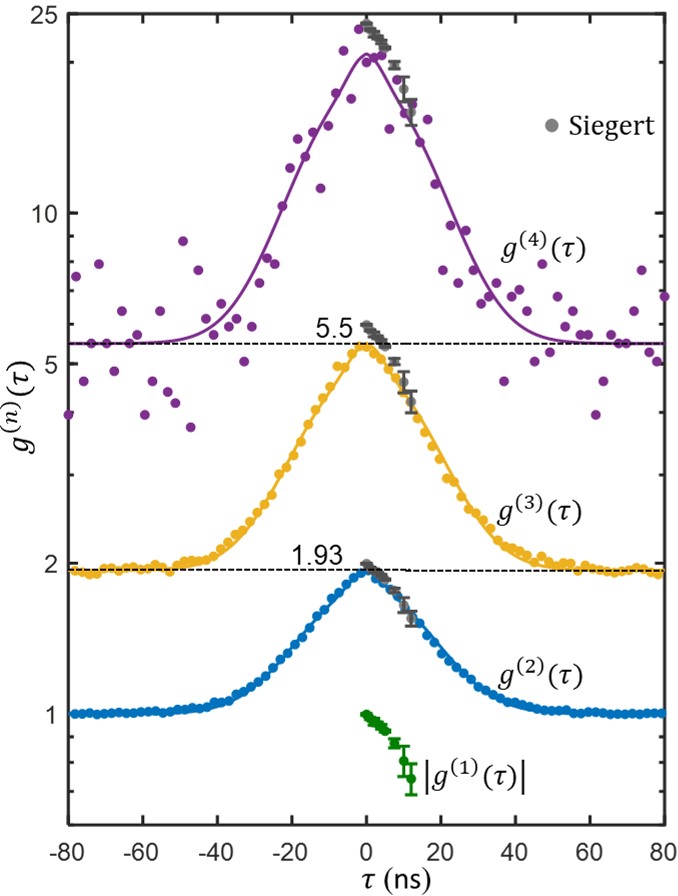}
	\caption{Measured correlation functions $g^{(n)}(\tau)$ of light scattered from a resonantly excited warm vapor of $^{87}\text{Rb}$ atoms, for $n=1,2,3,4$. The measured $g^{(n)}(0)$ values and the estimated temporal correlation widths confirm the chaotic nature of the scattered light, with a temporal coherence set by the Doppler broadening. The solid lines correspond to the theoretical correlation functions for chaotic light from atoms~\cite{dussaux2016temporal}, using generalized Siegert relations and including residual noise in form of higher-order $\beta$ parameters~\cite{Lemieux1999}. The grey points are evaluated directly by plugging the measured $|g^{(1)}(\tau)|$ into the ideal generalized Siegert relations~(\ref{eq:siegert_n}). Note that the limiting values of $g^{(n)}(\tau_1=0,...,\tau_{n-1}\rightarrow\infty)$ are given by the lower-order normalized correlation $g^{(n-1)}(\tau_1=0,...,\tau_{n-2}=0)$, and are affected by the residual Poissonian noise or multimodal character of detected light. 
 } 
	\label{fig:higher Siegert}
\end{figure}

The measured second-order correlation is plotted in Fig.~\ref{fig:higher Siegert}. The experimental data are fitted using a simple model considering the Doppler broadening and multiply-scattered photons in warm atomic vapor~\cite{dussaux2016temporal}, with an estimated single scattering probability of 91.3$\pm 0.7~\%$. The measured $g^{(2)}(0)=1.93\pm 0.02$ is close to the value of $2$ of chaotic light, and the FWHM of $28\pm 1$~ns corresponds to the Doppler broadening of the single-scattered photons at the chosen observation angle. The measured interference visibility in the Michelson interferometric setup allows us to evaluate $|g^{(1)}(\tau)|$ for various time delays $\tau$. Comparing $1+|\gtau|^2$ with $\ggtau$, the validity of the Siegert relation~\eqref{eq:Siegert_simple} is confirmed for this setup as well -- see Fig.~\ref{fig:higher Siegert}. 



Moving to higher order correlations, the normalized $n$-th order intensity correlation function is defined as
\begin{equation}
g^{(n)}(\tau_1,...,\tau_{n-1})=
\frac{\langle I_1(t)...I_n(t+\tau_{n-1})\rangle}{\langle I_1(t)\rangle...\langle I_n(t)\rangle},
\label{eq1}
\end{equation}
where $I_i(t+\tau_{i-1})=E_i^*(t+\tau_{i-1})E_i(t+\tau_{i-1})$ is the intensity detected by $i$-th detector~\cite{mandel1995optical}. The maximum value of $g^{(n)}(0)$ for chaotic light is given by $n!$~\cite{Glauber:1963b, wick1950evaluation,stevens2010high}. As for the case of $\ggtau$, for chaotic light the higher-order correlations can be expressed as a function of the $\gtau$ function. According to Isserlis' theorem~\cite{Isserlis_1918}, in the particular simplified case where we set $\tau_n=0$ for any $n>1$, the higher order correlation functions read
\begin{eqnarray}
g^{(n)}(\tau)=(n-1)!\big[1+(n-1)|g^{(1)}(\tau)|^2\big].
\label{eq:siegert_n}
\end{eqnarray}
Here, without loss of generality, $\tau$ represents the tunable delay between any two chosen detectors in the symmetric multiplexing setup, where all other mutual temporal delays are equal.

The measurements presented in Fig.~\ref{fig:higher Siegert} exhibit bunching for the normalized three-photon and four-photon correlations, with $g^{(3)}(0)=5.5\pm 0.03$ and $g^{(4)}(0)=20.8\pm 1.3$. Note that the signal-to-noise ratio decreases with the increasing correlation order $n$, evaluated over the same total measurement period. This is due to the exponential dependence of the corresponding mean probability of the $n$-fold coincidence on the overall photon detection efficiency,~$\sim\eta^n$, which is critical in detecting higher order photon correlations in the high-loss detection limit of free-space atomic light sources. 
Nevertheless, the measured correlation functions present a good agreement with the generalized Siegert relation~(\ref{eq:siegert_n}), consistently with a chaotic nature for the scattered light -- the ideal values being $g^{(3)}(0)=6$ and $g^{(4)}(0)=24$. Furthermore, the temporal profile of the different correlation functions are very similar, as expected from a chaotic behavior, and within evaluated error bars~\cite{liang2018observation}. The minor deviations observed when comparing the $g^{(n)}$ curves with the values predicted from the generalized Siegert relation~(\ref{eq:siegert_n}) and from the $|\gtau|$ measurement can be accounted for by simulating a model composed of an ideal chaotic light statistics with a small contribution of Poissonian noise: The relative probability of $0.04$ of the latter corresponds to the sum of all residual noise sources, such as detector background counts, leakage of the excitation laser to detectors, and temporal multi-modeness resulting from the finite resolution of the employed single-photon counting modules, particularly relevant for the large-bandwidth multiple-scattered photon contribution. Alternatively, this discrepancy can be conveniently incorporated in the higher-order $\beta$ parameters~\cite{Lemieux1999}. The fit of the generalized Siegert equation including residual multimodal effects as discussed in Ref.~\cite{Lemieux1999} leads to $\beta_1=0.93\pm 0.02, \beta_2=0.84\pm 0.02$, and $\beta_3=0.8\pm 0.04$. These values follow approximately the predicted scaling $\beta_n=\beta_1^n$ for uncorrelated chaotic light
modes~\cite{Lemieux1999}. The sensitivity to such deviations, including the estimation of the multi-modal character of the light field or its corruption by diverse noise sources, are substantially enhanced when measuring higher-order correlations: Many-photon correlations thus can be seen as a sensitive tool to probe and characterize small fluctuations beyond chaotic light~\cite{zhang2005correlated,valencia2005two,weedbrook2012continuous}.


\section{Conclusion}

We have investigated the chaotic nature of the light scattered from two complementary atomic setups, a cold and a warm one. In the former setup, an important benefit is the simultaneous measurement of $\gtau$ and $\ggtau$ functions within a single experiment, effectively circumventing potential experimental drift or fluctuations that would arise if conducted separately.  Additionally, this method does not require any moving components, nor precise control over different paths at scales smaller than the light wavelength. However, this technique relies on the LO being well under control in terms of phase noise, as compared to the collected scattered field. In the second setup, we have access to correlation functions up to the 4th order. While it is achievable when correlations are continuously measured, this is more challenging in a cold atom setup due to its limited duty cycle. The probe pulse in such setups lasts about 20 to 100~$\mu$s compared to a total cycle duration of about 100\,ms, significantly reducing the signal-to-noise ratio over the entire experimental duration. Finally, the Michelson interferometer offers a direct access to the~$|\gtau|$ function by measuring the fringe contrast, and higher resolution for small delays~$\tau$ compared to $\ggtau$ which can be impacted by the finite temporal resolution of employed single-photon detectors. This becomes particularly relevant for the measurements with hot atoms where a large bandwidth of multiple-scattered photon contributions imposes limits on detectors' temporal resolution on the nanosecond scale. Yet, the price to pay is a more challenging implementation for large $\tau$ for which the overall path fluctuations can significantly affect the interference contrast.

The two experimental setups are also complementary in terms of the physical process responsible for the chaotic nature of the scattered light. In the cold atom setup, the random nature of the emission stems from the quantum randomness of spontaneous emission, whereas in the warm vapor experiment it results from temperature-induced Doppler broadening. Both setups exhibit temporal correlations, at all measured orders, consistent with chaotic light. In particular, the generalized Siegert relation, which establishes a relation between the first-order correlation $\gtau$ and the higher-order ones, is satisfied in both.

The minor deviations from this relation can be attributed to technical artefacts such as intensity fluctuations of the probe, finite probe coherence, existence of a static field~\cite{Ferioli2023}, or beyond-single-mode detection (in terms of polarization or spatial modes). In this context, the higher sensitivity of higher-order correlations to these sources of noise make them a tool of choice to investigate the underlying mechanisms. Note that even without such artifacts, the Siegert relations may not be satisfied. For example, other properties related to the scattering process, such as intermittent dynamics, maybe affect differently the field-field and photon-photon correlations\,\cite{Lemieux2001}. Finally, while the moments may obey the Siegert relations, deviations at nonzero delays can occur when the relaxation and coherence times are different\,\cite{Lemieux2001}, or in the presence of emitter correlations giving rise to sub- and super-radiant coupling in partially coherent nanolasers for instance\,\cite{Drechsler_2022}. Hence, beyond the emblematic Siegert relation which applies to the broad class of light whose field exhibit Gaussian statistics, the study of temporal correlations at different orders remains an important tool to characterize the nature of the photon emission and the scattering processes behind it. One can cite, for example, the characterization of the laser threshold\,\cite{Chow_2014}, or the tunability of light states with an increased second-order coherence, yet a low first-order coherence as in quantum dot superluminescent diodes~\cite{Blazek_2011}.



\acknowledgments
The authors thank William Guerin for his careful reading for his careful reading and comments on the manuscript. M.\,H. and R.\,K. acknowledge funding from the French National Research Agency (ANR) (QuaCor ANR19-CE47-0014-01), the European Unions Horizon 2020 research and innovation program in the framework of Marie Skodowska-Curie HALT project (grant agreement No 823937) and the Doeblin fundation.  M.\,H. and R.\,B. has been supported by the French government, through the UCA J.E.D.I. Investments in the Future project managed by the ANR with the reference number ANR-15-IDEX-01. R.K. received support from the European project ANDLICA, ERC Advanced Grant Agreement No. 832219.
R.\,K., R.\,B. and M.\,H. received support from the project STIC-AmSud (Ph879-17/CAPES 88887.521971/2020-00) and with J.\,C.\,C.\,C. from CAPES-COFECUB (CAPES 88887.711967/2022-00). C. R.\,B. benefited from Grants from S\~ao Paulo Research Foundation (FAPESP, Grants Nos. 2018/15554-5, 2019/13143-0, 2022/00209-6, 2023/07100-2 and 2023/03300-7) and from the National Council for Scientific and Technological Development (CNPq, Grant No.\,313632/2023-5). S.~J., J.~M., and L.~S. are grateful to MEYS of Czech Rep. for support under grant agreement No. 731473 and, with R.\,K. and M.\,H., from the QUANTERA ERA-NET cofund in quantum technologies within the European Union’s Horizon 2020 Programme (PACE-IN, 8C20004, ANR19-QUAN-003-01).

\bibliographystyle{eplbib}
\bibliography{Biblio}

\begin{thebibliography}{10}
\expandafter\ifx\csname url\endcsname\relax\def\url#1{\texttt{#1}}\fi

\bibitem{Glauber:1963a}
\Name{Glauber R.~J.} \REVIEW{Phys. Rev.}{130}{1963}{2529}.

\bibitem{Martienssen_1964}
\Name{Martienssen W. \and Spiller E.} \REVIEW{American Journal of
  Physics}{32}{1964}{919}.

\bibitem{Lassegues2023}
\Name{Lass\`egues P., Biscassi M. A.~F., Morisse M., Cidrim A., Dias P. G.~S.,
  Eneriz H., Teixeira R.~C., Kaiser R., Bachelard R. \and Hugbart M.}
  \REVIEW{Phys. Rev. A}{108}{2023}{042214}.

\bibitem{Goodman:1976}
\Name{Goodman J.~W.} \REVIEW{J. Opt. Soc. Am.}{66}{1976}{1145}.

\bibitem{Glauber:1963b}
\Name{Glauber R.~J.} \REVIEW{Phys. Rev.}{131}{1963}{2766}.

\bibitem{Lemieux1999}
\Name{Lemieux P.-A. \and Durian D.~J.} \REVIEW{Journal of the Optical Society
  of America A}{16}{1999}{1651}.

\bibitem{titulaer1965correlation}
\Name{Titulaer U. \and Glauber R.} \REVIEW{Physical Review}{140}{1965}{B676}.

\bibitem{sperling2015uncovering}
\Name{Sperling J., Bohmann M., Vogel W., Harder G., Brecht B., Ansari V. \and
  Silberhorn C.} \REVIEW{Physical Review Letters}{115}{2015}{023601}.

\bibitem{laiho2022measuring}
\Name{Laiho K., Dirmeier T., Schmidt M., Reitzenstein S. \and Marquardt C.}
  \REVIEW{Physics Letters A}{435}{2022}{128059}.

\bibitem{dall2013ideal}
\Name{Dall R., Manning A., Hodgman S., RuGway W., Kheruntsyan K.~V. \and
  Truscott A.} \REVIEW{Nature Physics}{9}{2013}{341}.

\bibitem{liang2018observation}
\Name{Liang Q.-Y., Venkatramani A.~V., Cantu S.~H., Nicholson T.~L., Gullans
  M.~J., Gorshkov A.~V., Thompson J.~D., Chin C., Lukin M.~D. \and Vuleti{\'c}
  V.} \REVIEW{Science}{359}{2018}{783}.

\bibitem{patil2022measuring}
\Name{Patil Y.~S., Yu J., Frazier S., Wang Y., Johnson K., Fox J., Reichel J.
  \and Harris J.~G.} \REVIEW{Physical Review Letters}{128}{2022}{183601}.

\bibitem{assmann2009higher}
\Name{A{\ss}mann M., Veit F., Bayer M., van~der Poel M. \and Hvam J.~M.}
  \REVIEW{Science}{325}{2009}{297}.

\bibitem{Siegert:1943}
\Name{Siegert A. J.~F.} Report: Radiation Laboratory Massachusetts Insitute of
  Technology (1943).

\bibitem{Loudon:book}
\Name{Loudon R.} \Book{The quantum theory of light} (Oxford Science
  Publications) 1973.

\bibitem{Ortiz_2019}
\Name{Ortiz L., Teixeira R.~C., Eloy A., Ferreira D., Kaiser R., Bachelard R.
  \and Fouch\'e M.} \REVIEW{New Journal of Physics}{21}{2019}{093019}.

\bibitem{Lassegues_2022}
\Name{Lass{\`e}gues P., Biscassi M. A.~F., Morisse M., Cidrim A., Matthews N.,
  Labeyrie G., Rivet J.-P., Vakili F., Kaiser R., Guerin W., Bachelard R. \and
  Hugbart M.} \REVIEW{The European Physical Journal D}{76}{2022}{246}.

\bibitem{HBT:1956a}
\Name{{Hanbury~Brown} R. \and Twiss R.~Q.} \REVIEW{Nature}{177}{1956}{27}.

\bibitem{Steudle2012}
\Name{Steudle G.~A., Schietinger S., H\"ockel D., Dorenbos S.~N., Zadeh I.~E.,
  Zwiller V. \and Benson O.} \REVIEW{Phys. Rev. A}{86}{2012}{053814}.

\bibitem{Ferreira2020}
\Name{Ferreira D., Bachelard R., Guerin W., Kaiser R. \and Fouch{\'{e}} M.}
  \REVIEW{American Journal of Physics}{88}{2020}{831}.

\bibitem{Mollow_1969}
\Name{Mollow B.~R.} \REVIEW{Phys. Rev.}{188}{1969}{1969}.

\bibitem{Goodman:StatOptics}
\Name{Goodman J.} \Book{Statistical Optics} Wiley Series in Pure and Applied
  Optics (Wiley) 2015.

\bibitem{mandel1995optical}
\Name{Mandel L. \and Wolf E.} \Book{Optical coherence and quantum optics}
  (Cambridge university press) 1995.

\bibitem{Isserlis_1918}
\Name{Isserlis L.} \REVIEW{Biometrika}{12}{1918}{134}.

\bibitem{wick1950evaluation}
\Name{Wick G.-C.} \REVIEW{Physical review}{80}{1950}{268}.

\bibitem{Ficek_2005}
\Name{Ficek Z. \and Swain S.} \Book{Quantum interference and coherence: theory
  and experiments} Vol. 100 (Springer Science \& Business Media) 2005.

\bibitem{Marconi_2018}
\Name{Marconi M., Javaloyes J., Hamel P., Raineri F., Levenson A. \and
  Yacomotti A.~M.} \REVIEW{Phys. Rev. X}{8}{2018}{011013}.

\bibitem{Wolf1988}
\Name{Wolf P., Maret G., Akkermans E. \and Maynard R.} \REVIEW{Journal de
  Physique}{49}{1988}{63}.

\bibitem{Ferioli2023}
\Name{Ferioli G., Pancaldi S., Glicenstein A., Clement D., Browaeys A. \and
  Ferrier-Barbut I.} \Book{Non-gaussian correlations in the steady-state of
  driven-dissipative clouds of two-level atoms} (2023).

\bibitem{Voigt1994}
\Name{Voigt H. \and Hess S.} \REVIEW{Physica A: Statistical Mechanics and its
  Applications}{202}{1994}{145}.

\bibitem{Kovalenko_2023}
\Name{Kovalenko A., Babjak D., Le\v{s}und\'{a}k A., Podhora L., Lachman L.,
  Ob\v{s}il P., Pham T., \v{C}\'{i}p O., Filip R. \and Slodi\v{c}ka L.}
  \REVIEW{Optica}{10}{2023}{456}.

\bibitem{mika2018generation}
\Name{Mika J., Podhora L., Lachman L., Ob{\v{s}}il P., Hlou{\v{s}}ek J.,
  Je{\v{z}}ek M., Filip R. \and Slodi{\v{c}}ka L.} \REVIEW{New Journal of
  Physics}{20}{2018}{093002}.

\bibitem{dussaux2016temporal}
\Name{Dussaux A., de~Silans T.~P., Guerin W., Alibart O., Tanzilli S., Vakili
  F. \and Kaiser R.} \REVIEW{Physical Review A}{93}{2016}{043826}.

\bibitem{Eloy_2018}
\Name{Eloy A., Yao Z., Bachelard R., Guerin W., Fouch\'e M. \and Kaiser R.}
  \REVIEW{Phys. Rev. A}{97}{2018}{013810}.

\bibitem{oppel2012superresolving}
\Name{Oppel S., B{\"u}ttner T., Kok P. \and von Zanthier J.} \REVIEW{Physical
  review letters}{109}{2012}{233603}.

\bibitem{avenhaus2010accessing}
\Name{Avenhaus M., Laiho K., Chekhova M. \and Silberhorn C.} \REVIEW{Physical
  review letters}{104}{2010}{063602}.

\bibitem{qi2018multiphoton}
\Name{Qi L., Manceau M., Cavanna A., Gumpert F., Carbone L., Vittorio M.~d.,
  Bramati A., Giacobino E., Lachman L., Filip R. \etal} \REVIEW{New Journal of
  Physics}{20}{2018}{073013}.

\bibitem{gol2001picosecond}
\Name{Gol’Tsman G., Okunev O., Chulkova G., Lipatov A., Semenov A., Smirnov
  K., Voronov B., Dzardanov A., Williams C. \and Sobolewski R.} \REVIEW{Applied
  physics letters}{79}{2001}{705}.

\bibitem{stevens2010high}
\Name{Stevens M.~J., Baek B., Dauler E.~A., Kerman A.~J., Molnar R.~J.,
  Hamilton S.~A., Berggren K.~K., Mirin R.~P. \and Nam S.~W.} \REVIEW{Optics
  Express}{18}{2010}{1430}.

\bibitem{zhang2005correlated}
\Name{Zhang D., Zhai Y.-H., Wu L.-A. \and Chen X.-H.} \REVIEW{Optics
  letters}{30}{2005}{2354}.

\bibitem{valencia2005two}
\Name{Valencia A., Scarcelli G., D’Angelo M. \and Shih Y.} \REVIEW{Physical
  review letters}{94}{2005}{063601}.

\bibitem{weedbrook2012continuous}
\Name{Weedbrook C., Pirandola S. \and Ralph T.~C.} \REVIEW{Physical Review
  A}{86}{2012}{022318}.

\bibitem{Lemieux2001}
\Name{Lemieux P.-A. \and Durian D.~J.} \REVIEW{Appl. Opt.}{40}{2001}{3984}.

\bibitem{Drechsler_2022}
\Name{Drechsler M., Lohof F. \and Gies C.} \REVIEW{Applied Physics
  Letters}{120}{2022}{221104}.
\newline\url{https://doi.org/10.1063/5.0094698}

\bibitem{Chow_2014}
\Name{Chow W.~W., Jahnke F. \and Gies C.} \REVIEW{Light: Science \&
  Applications}{3}{2014}{e201}.

\bibitem{Blazek_2011}
\Name{Blazek M. \and Els\"a\ss{}er W.} \REVIEW{Phys. Rev. A}{84}{2011}{063840}.

\end{thebibliography}

\end{document}